\documentclass[aps,prl,twocolumn,groupedaddress,showpacs,amssymb,amsmath,letterpaper]{revtex4}
\usepackage{graphicx}
\usepackage{psfrag} 
\usepackage{booktabs}
\usepackage{subfigure}
\usepackage[final,dvips]{epsfig}
\usepackage{color}
\usepackage{dcolumn}
\usepackage{bm}
\usepackage{array}
\usepackage{amssymb}
\usepackage{mathptmx}
\newcommand{\bi}{\begin{itemize}}
\newcommand{\ei}{\end{itemize}}
\newcommand{\be}{\begin{equation}}
\newcommand{\ee}{\end{equation}}
\newcommand{\ba}{\begin{eqnarray}}
\newcommand{\ea}{\end{eqnarray}}

\newcommand{\COMMENTED}[1]{}

\begin{document}

\title{Spin and charge order in doped Hubbard model: long-wavelength collective modes}

\author{Chia-Chen Chang}

\author{Shiwei Zhang}

\affiliation{Department of Physics, College of William and Mary, Williamsburg, VA 23187}

\begin{abstract}   

Determining the ground state properties of the two-dimensional
Hubbard model has remained an outstanding problem.
Applying recent advances in constrained path  auxiliary-field
quantum Monte Carlo techniques and simulating 
large rectangular periodic lattices, 
we calculate the long-range spin and charge correlations in
the ground state as a function of doping.
At intermediate interaction strengths, an incommensurate spin
density wave (SDW) state is found, with antiferromagnetic order and
essentially homogeneous charge correlation.
The wavelength of the collective mode 
decreases with doping, as does its magnitude.
The SDW order vanishes beyond a critical doping.
As the interaction is increased, the holes go from a wave-like to a
particle-like state, and charge ordering develops which
eventually evolves into stripe-like states.

\end{abstract}


\maketitle

The Hubbard model \cite{Hubbard} is one of the simplest and most fundamental
models for quantum many-body systems.
Since the discovery of high-$T_c$ superconductors in 1986,
the two-dimensional (2D) repulsive Hubbard model has received great interest,
as a potential model for describing the essential physics
of the copper-oxygen plane in cuprates \cite{Anderson1987}.
Thanks to intensive analytic and numerical investigations,
some aspects of its phase diagram have been understood \cite{HubbardReview}.
For example, at half-filling (one electron per lattice site),
the system has long-range antiferromagnetic (AF) order 
\cite{Hirsch,White1989,Furukawa1992,Varney2009}.
However, many basic issues still remain unknown or controversial.
For instance, what happens to the AF order when the system is doped? 
This is important not only for understanding the magnetic properties,
which the Hubbard model was originally designed to describe,
but also in the context of high-$T_c$, which shares the same 
parameter regime and is believed to be closely related to
AF fluctuations.

The difficulty in treating the Hubbard model underscores a more general 
theme, namely the challenge of accurate treatment of strongly correlated 
systems.
Recent experimental work \cite{FermionOpticalLattice} 
with cold fermionic atoms in optical lattices offers a promising new 
avenue --- potentially direct simulations of Hubbard-like models with 
``lattice emulators'' \cite{Emulator}.
We believe this increases, rather
than decreases, the demand on ``traditional'' numerical simulations. 
Although all numerical methods 
have their limitations, high quality data on the Hubbard model 
will provide guidance and allow direct comparison 
with experiments, thereby creating
a new level of synergy to tackle the problem of strong electron 
correlations.

There are many earlier investigations of the spin and charge correlations
in the 2D Hubbard model 
\cite{HubbardReview,Hirsch,Furukawa1992,Varney2009,Imada1989,White1989,Moreo1990,Lin1991,Fano1992,Cosentini1998,Becca2000,
Stripe,Maier2005,Aichhorn2007,Aimi2007},
but reaching large system sizes 
with sufficient accuracy has been a challenge.
Recently, we have calculated the equation of state 
from low to intermediate interaction strengths on periodic
lattices of up to $\sim 16\times 16$ \cite{Chang2008}.
The constrained path Monte Carlo (CPMC) method \cite{CPMC,Zhang2003}
was generalized to incorporate 
a boundary condition integration technique \cite{TABC},
which removed short-range finite-size effects. 
It was found that, immediately upon doping, 
the thermodynamic stability condition is violated. 
This implied the existence of a spatially inhomogeneous phase.
In the absence of long-range collective modes, the results
are an accurate representation of the thermodynamic limit,
and the instability would indicate phase separation.
On the other hand, if
long-range collective modes existed whose characteristic
length exceeds the size of the super cell ($\sim 16$),
they would not be fully captured in the simulations.
The nature of the AF fluctuation in the doped Hubbard model thus remains 
to be resolved.

Here we address the question,
by employing recent algorithmic advances 
\cite{Chang2008,Kwee2008,Unpublished2009}
in CPMC and simulating  
rectangular supercells on parallel computers. 
Much larger linear dimension ($128$) is reached than in previous studies
($\sim 16$),
and detailed measurements are obtained
of the spin-spin and charge-charge correlations in
the ground state, at intermediate interaction strengths where 
our method is very accurate.
Our results show that
long wavelength collective modes of incommensurate spin
density wave (SDW) states appear as the system is doped, 
with AF spin order but
essentially homogeneous charge correlation. 
Charge correlation develops as the interaction is further increased.
We quantify the nature of such states, and 
discuss how they relate to
the ``stripe'' states at large interactions.

The Hamiltonian for the single-band Hubbard model is 
\begin{eqnarray}
  H = H_1+H_2 
   &=& -t \sum_{\mathbf j, \bm\delta,\sigma} 
        c_{\mathbf j,\sigma}^\dagger c_{\mathbf j+\bm\delta,\sigma} 
      + U\sum_{\mathbf j} n_{\mathbf j\uparrow} n_{\mathbf j\downarrow},
\label{eq:hamiltonian}
\end{eqnarray}
where $c_{\mathbf{j},\sigma}^\dagger$ ($c_{\mathbf{j},\sigma}$) creates 
(annihilates) an electron with spin $\sigma$ ($\sigma=\uparrow, \downarrow$) 
at lattice site $\mathbf{j}$,  and $\bm\delta$ connects all possible 
nearest-neighbor sites.
The supercell has $N=L_x\times L_y$ sites. The density is 
$n\equiv (N_\uparrow+N_\downarrow)/N$, 
where $N_\sigma$ is the number of electrons with spin $\sigma$;
doping is $h\equiv1-n$. 
We implement twist-averaged boundary conditions \cite{TABC}, 
under which the wave function gains 
a phase when electrons hop around lattice boundaries:
$
  \Psi(\ldots,\mathbf{r}_j+\mathbf{L},\ldots) = e^{i\widehat{\mathbf{L}}\cdot{\bm\Theta}}
  \Psi(\ldots,\mathbf{r}_j,\ldots),
$
where $\widehat{\mathbf{L}}$ is the unit vector along $\mathbf{L}$,
and ${\bm\Theta}=(\theta_x,\theta_y)$ are random twist angles 
over which we average.

The generalized CPMC method \cite{CPMC,Zhang2003,Chang2008,Unpublished2009}
used here obtains the many-body ground
state by repeated projections with $e^{-\tau H}$ ($\tau$ is the projection time step),
as in standard quantum Monte Carlo (QMC). The two-body part,
$e^{-\tau H_2}$, is decoupled via the Hubbard-Stratonovich transformation
into a 
sum over one-body projectors in Ising fields \cite{Hirsch}.
The projection is then realized efficiently by importance-sampled
random walks with non-orthogonal Slater determinants (SDs), where the one-body 
projectors propagate one SD into another, 
and the many-dimensional
sum over Ising fields is performed by Monte Carlo.
The usual fermion sign/phase problem is controlled approximately by a
{\em global\/} phase condition on the SDs. 
This is the only approximation in the method.
The basic idea of the approximation is as follows. 
The many-body ground state is given by 
$|\Psi_0\rangle = \sum_\phi w(\phi) |\phi\rangle$, where
$|\phi\rangle$ are SDs 
sampled by the QMC,
and their probability distribution will give $w(\phi)$ ($>0$).
Because the Schr\"odinger equation
is linear, $|\Psi_0\rangle$ is degenerate with $-|\Psi_0\rangle$.
A trivial effect in a deterministic representation,
this can cause the determinants $|\phi\rangle$, in a random walk,
to move back and forth 
between the two sets of solutions.
In a simulation, precisely 
when a $|\phi\rangle$ turns from one to the other can not be detected, 
because the continuous stochastic evolution of the 
orbitals can lead to an exchange without any two orbitals ever overlapping.
This is the sign problem.
We use a trial wave function $|\Psi_T\rangle$ to make the detection, by
requiring $\langle\Psi_T|\phi\rangle>0$.
Because each $|\phi\rangle$ is a full many-electron wave function,
the sign of its overlap with a $|\Psi_T\rangle$ is expected to be  
quite insensitive to the details of $|\Psi_T\rangle$.

In extensive benchmarks in 
Hubbard models \cite{CPMC,Purwanto2004,Chang2008} as well as in
atoms, molecules, and
solids \cite{Benchmark}, 
this general framework
has demonstrated accuracy equaling or surpassing
the most accurate (non-exponential scaling) many-body
computational methods available.
In the Hubbard model, 
the energy is typically within 
$< 0.5\%$ of the exact diagonalization results for $U=4t$ \cite{Chang2008}.
At half-filling where the approximation is 
the most severe (with free-electron $|\Psi_T\rangle$),
the method gives an energy per site of $-0.8559(4)$
for the infinite lattice, compared to the estimated exact result of $-0.8618(2)$
\cite{Becca2000,Sorella_priv}. 
(The method can be made exact at half-filling by removing the 
constraint \cite{CPMC}.)

\begin{figure}
  \psfrag{e}[0][0][1][0]{$\epsilon(n)-\epsilon_M(n)$}
  \psfrag{f}[0][0][1.5][0]{$\epsilon$}
  \includegraphics[width=0.30\textwidth, height=0.42\textwidth,angle=-90]{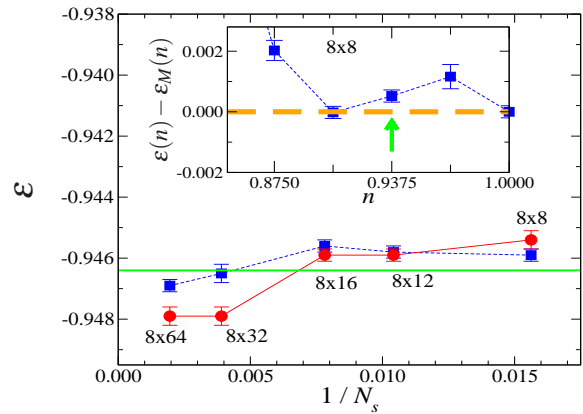} 
  \caption{
  Energy variations with linear dimension.
  The QMC ground state energy per site, $\epsilon$ (in units of $t$), 
  is shown for
  $U=4t$ for a sequence of supercell sizes $8\times L_y$.
  Blue squares are with an FE $|\Psi_T\rangle$, while 
  red circles are with a UHF $|\Psi_T\rangle$.
  The main graph is for $h=1/16$.
  The green line is determined by Maxwell construction, 
  which is illustrated in the inset.
  To magnify the vertical scale in the inset, a linear shift of $\epsilon_M(n)$
  has been applied, so 
  the tangent line is horizontal (dashed orange), and the EOS is plotted as
  $\epsilon(n)-\epsilon_M(n)$.
  }
  \label{fig:energy}
\end{figure}

In order to probe correlations at long range, we study rectangular 
supercells of $4\times L_y$, $8\times L_y$ and  $16\times L_y$.
The largest system size simulated in this work is $N\sim 1024$, using
$\mathcal{O}(1000)$ 
processors on the Cray XT4 supercomputer.
The first indication of a long wavelength collective mode is seen in 
the ground-state energy.
Figure~\ref{fig:energy} shows how the energy per site,
$\epsilon$, varies as $L_y$ is increased. 
Each energy has been averaged over 20-1000 $\Theta$-values, 
and all controllable QMC biases (e.g., Trotter and population size 
\cite{CPMC}) have been removed.
The error bars are estimated by combining statistical error and 
$\Theta$-point fluctuation. 
Twist-averaging eliminates kinetic energy finite-size 
effects (shell and lattice size) \cite{Chang2008}.
At $h=1/4$ for example, we have reconfirmed that 
the energy remained essentially constant as $L_y$ was varied from $8$ to 
$64$.

At $h=1/16$, the energies remain 
{\em above\/} the line from Maxwell construction 
up to $L_y\sim 16$,
where it shows a significant drop and falls {\em below\/} the line.
In the inset, the equation of state (EOS), $\epsilon(n)$ vs.~$n$, 
is shown for $8\times 8$. The EOS is concave for 
$n\in (n_c,1)$ \cite{Chang2008}.
The critical density $n_c$ is determined by the
Maxwell construction, which gives a phase separation line 
tangent to the EOS:
$\epsilon_M(n)=
h/h_c\epsilon(n_c)+(1-h/h_c)\epsilon(n=1)$.
The drop for $L_y>16$
indicates that the instability occurs only in smaller supercells, 
in which a state with long-range correlation is frustrated.

We use two different types of $|\Psi_T\rangle$, 
to help gauge the effect of the constraint. 
The first is the free-electron (FE) wave function, which 
is of course homogeneous with no long-range correlation. The second is the
unrestricted Hartree-Fock (UHF) solution, which has broken spatial symmetry 
and static long-range spin and charge order. 
We have verified that, in the paramagnetic phase below $n_c$, 
the two types lead to statistically indistinguishable QMC results. 
As seen in Fig.~\ref{fig:energy}, the same is true at $h=1/16$, 
{\em except for\/} large systems ($L_y> 16$),
where the UHF $|\Psi_T\rangle$ gives lower energy.
This is consistent with UHF being a better
wave function in a system where an SDW can develop, 
as further discussed below.

We calculate the spin-spin correlation function:
\be
\label{eq:CorrSpin}
C_s(\mathbf{r})=\frac{1}{N}\sum_{\mathbf{r}'}
\left\langle
\left(n_{\mathbf{r}+\mathbf{r}',\uparrow}-n_{\mathbf{r}+\mathbf{r}',\downarrow}\right)
\left(n_{\mathbf{r}',\uparrow}-n_{\mathbf{r}',\downarrow}\right)
\right\rangle,
\ee
which measures the correlation between two spins 
separated by a lattice vector
$\mathbf{r}\equiv (l_x,l_y)$. 
The corresponding structure factor is
$S_s(\mathbf{k})\equiv
\sum_\mathbf{r}\,e^{i\mathbf{k}\cdot\mathbf{r}}\,C_s(\mathbf{r})$.
Similarly, we calculate the charge-charge correlation 
$C_c({\mathbf r})$, defined by replacing ``$-$''
in Eq.~(\ref{eq:CorrSpin}) with  ``$+$'', and its
structure factor $S_c(\mathbf{k})$.

\begin{figure}
  \begin{center}
    \psfrag{a}[0][0][1][0]{$l_x$}
    \psfrag{b}[0][0][1][0]{$C_s^\prime(l_x,l_y)$}
    \psfrag{c}[0][0][1][0]{$l_x$}
    \psfrag{d}[0][0][1][0]{$l_y$}
    \psfrag{e}[0][0][1][0]{$C_s^\prime(l_x,l_y)$}
    \psfrag{f}[0][0][1][90]{$C_s(\mathbf{r})$}
    \psfrag{g}[0][0][1][90]{$C_s(\mathbf{r})$}
    \includegraphics[scale=0.415,angle=-90]{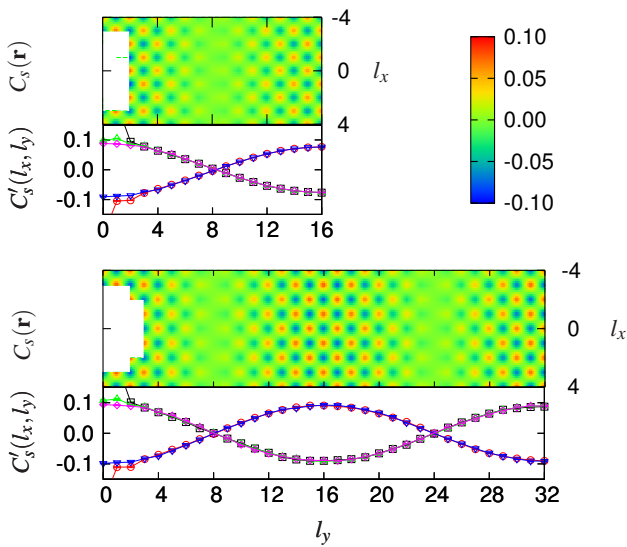}
    \caption{
    Spin-spin correlation function
    in $8\times 32$ 
    (top panel) and $8\times 64$ (bottom panel) lattices at $h=1/16$
    and $U=4t$.
    The results are averaged over 20 $\bm\Theta$-points. 
    UHF $|\Psi_T\rangle$'s are used.
    The upper part in each panel is a 3D 
    density plot (color theme in the upper right corner of the graph).
    The lower part of each panel shows the
    {\em staggered} correlation,
    with curves of different colors representing different $l_x$'s.
    Due to symmetry, only $l_y\in[0,L_y/2]$ is shown.
    The range of $C_s(\mathbf{r})$ is 
    restricted to $[-0.1,0.1]$, so the ``self-peak'' 
    near the origin is cut off.
    }
    \label{fig:spin-correlation}
  \end{center}
\end{figure}

The QMC results for $C_s(\mathbf{r})$ are shown
in Fig.~\ref{fig:spin-correlation}.
An AF correlation
is seen clearly in the density plots, in which the signs alternate
for near neighbors. 
The {\em staggered\/} correlation function:
$C_s^\prime(l_x,l_y)\equiv (-1)^{l_y}C_s(l_x,l_y)$,
is also plotted with statistical error bars.
The curves fall into two groups, for even and odd $l_x$, respectively.
With perfect AF correlation, 
each group would be a constant function of $l_y$. 
In these systems the AF correlation patterns are modulated by 
a wave along $l_y$. 
A $\pi$ phase-shift occurs at 
the nodes where $C_s^\prime(l_x,l_y)$ 
crosses zero.
The wave essentially doubles as $L_y$ is doubled, with  
comparable SDW amplitude. We see that the wavelength of the spin modulation 
is $\lambda \sim 32$ at this density, consistent with 
Fig.~\ref{fig:energy}, 
where the energy lowering only occurs at 
$L_y\gg 16$.

\begin{figure}
  \psfrag{C}[0][0][1.1][0]{$C_s^\prime(l_x,l_y)$}
  \psfrag{z}[0][0][1.1][0]{$l_y$}
  \includegraphics[scale=0.35,angle=-90]{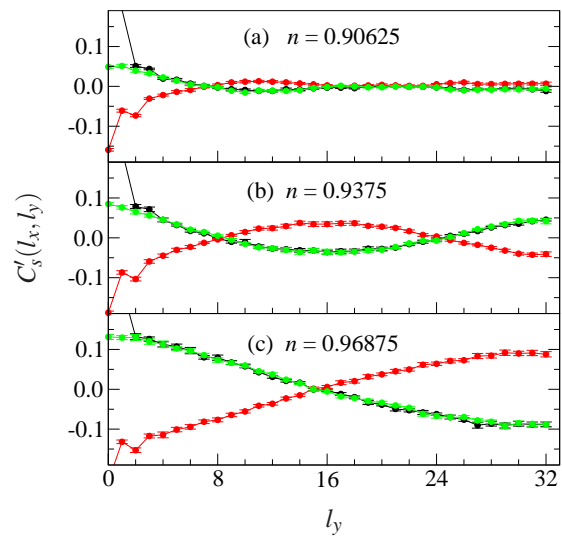} 
  \caption{
  Doping dependence of the long wavelength incommensurate SDW state.
  The staggered spin-spin correlation function $C_s^\prime(l_x,l_y)$
  is plotted vs.~$l_y$, at three different densities.
  Calculations are done using {\em free-electron\/} (FE) $|\Psi_T\rangle$.
  The system is a $4\times 64$ supercell, with $U=4t$ and
  ${\bm\Theta}/\pi=(-0.8410,-0.9198)$.
  Colors label different value of $l_x$'s as in 
  Fig.~\ref{fig:spin-correlation}.
  As doping is increased, the wavelength of the modulating wave
  decreases, as does the amplitude of the SDW.
  }
  \label{fig:doping}
\end{figure}

We next study the spin-spin correlation as a function of doping. 
Calculations are done at three densities,
$h=3/32$, $1/16$, and $1/32$, respectively, for a 
$4\times 64$ lattice, at $U=4t$.
The staggered spin-spin correlation function 
$C_s^\prime(\mathbf{r})$ 
are shown in Fig.~\ref{fig:doping}, where the
modulation and $\pi$ phase shifts are clearly seen.
The wavelength of the modulation 
decreases with doping, beginning at half-filling ($h=0$) where the
wavelength is $\lambda=\infty$. (Quantitatively, our data is consistent with 
the wavelength 
being inversely proportional to $h$, although statistical 
error bars on $\lambda$ are large.)
The strength of $C_s^\prime(\mathbf{r})$ in the 
incommensurate SDW state also decreases with doping, and appears to 
vanish at a critical value of $h_c\sim 0.15\pm 0.05$, where 
the system turns into a paramagnetic liquid.

The results in Fig.~\ref{fig:spin-correlation} 
were obtained using UHF $|\Psi_T\rangle$, while 
those in Fig.~\ref{fig:doping} were from 
FE $|\Psi_T\rangle$. The consistency between them is reassuring.
To generate the UHF $|\Psi_T\rangle$, 
we used the minimum $U$ for which a UHF solution exists
($U\sim 1.3$-$1.5t$ for $h=1/16$ and $\sim 3.5t$ for $h=1/4$), 
in order to minimize the effect of broken translational invariance.
A weak static long-range order is present in the UHF solution.
For example, in the $8\times 64$ system, 
$C_s$ calculated from the UHF $|\Psi_T\rangle$ itself was 
$\mathcal{O}(5 \times10^{-4})$.
We see that this was enhanced by a factor of $ 200$ in the QMC, to 
$\mathcal{O}(0.1)$.
On the other hand, at large distance the variation in the
UHF charge-charge correlation 
$C_c$ was $\mathcal{O}(1.5 \times10^{-4})$,
which remained $\mathcal{O}(10^{-4})$
in the QMC (zero within error bars). 
With the FE $|\Psi_T\rangle$, the SDW structure in 
Fig.~\ref{fig:doping} 
{\em emerged spontaneously\/}.
QMC results with UHF $|\Psi_T\rangle$ always showed long-range order, 
while we do find variations when
the FE $|\Psi_T\rangle$ is used: for some
$\bm\Theta$'s no long-range SDW order is seen, only ``short''-range
incommensurate AF correlations.
In such cases, the calculated QMC energy
tends to be higher than that
using UHF $|\Psi_T\rangle$.                       
In contrast, when
a long-range SDW is also seen with the FE
$|\Psi_T\rangle$, the QMC energies are more consistent
between the two types of $|\Psi_T\rangle$'s.
We interpret this as the system favoring long-range order.
The rectangular supercells break the symmetry between the $x$- and $y$-
directions. In the
thermodynamic limit a combination of the
linear SDWs may be present.

\begin{figure}
  \psfrag{a}[0][0][1.1][0]{$S_s(\pi,k_y)$}
  \psfrag{b}[0][0][1.1][0]{$k_y$}
  \psfrag{c}[0][0][1.1][0]{$k_y$}
  \psfrag{d}[0][0][1.1][0]{$S_c(0,k_y)$}
  \includegraphics[scale=0.3,angle=-90]{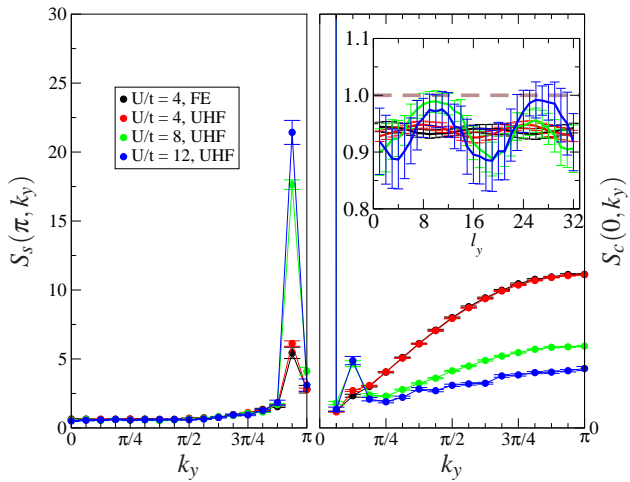} 
  \caption{
  Spin-spin and charge-charge 
  structure factors as a function of interaction strength $U$.
  Results are shown for a $4\times 32$ supercell, with
  $\Theta/\pi=(-0.8410,-0.9198)$ and at doping $h=1/16$.
  FE and UHF indicate the type of $|\Psi_T\rangle$ used in the QMC 
  calculation. 
  The inset in the right panel shows real space density profile 
  with the same color coding as in the main figures.  
  $S_s(\mathbf{k})$ has a peak at 
  $\mathbf{k}=(\pi,15\pi/16)$, with the peak value growing rapidly with
  increasing $U$.
  $S_c(\mathbf{k})$, which has no feature at $U=4t$, shows a corresponding 
  peak at $\mathbf{k}=(0,\pi/8)$ for large $U$. 
  The real-space density is a constant at 
  $U=4t$, but develops periodic modulation at larger $U$.
  }
  \label{fig:large-U}
\end{figure}

The spin correlation discussed above ($U=4t$) 
is not accompanied by charge inhomogeneities.
The effect of stronger interactions is examined in Fig.~\ref{fig:large-U},
which displays results 
for $U/t=4$, $8$ and $12$, with doping of $h=1/16$ in a
$4\times 32$ lattice. 
The spin structure factor 
$S_s(\mathbf{k})$ is plotted along the line cut $\mathbf{k}=(\pi,k_y)$.
At $U=4t$ a pronounced peak can already be seen 
at $(\pi,15\pi/16)$,
consistent with the spin-spin correlation in Fig.~\ref{fig:spin-correlation}.
(The split of the $S_s(\mathbf{k})$ peak with doping
had also been observed in earlier simulations 
\cite{HubbardReview,Imada1989,Moreo1990,Furukawa1992}.)
As $U$ is increased, the peak value increases rapidly.
The charge structure factor is plotted along $(0,k_y)$.
Except for the trivial peak at the origin, $S_c(\mathbf{k})$ is 
broad with little features at $U=4t$.
Above $U=8t$, a 
peak appears at $(0,\pi/8)$, indicating the development of a 
charge-charge correlation.

In the inset of Fig.~\ref{fig:large-U}, the real-space
density profile is
plotted along $(0,l_y)$. 
At $U=4t$, results using FE and
UHF (generated with $U= 1.4t$)
trial wave functions both give a constant density. 
At larger $U$, the same 
UHF $|\Psi_T\rangle$ 
(from $U=1.4t$) turned out to be sufficient
to ``pin'' the many-body solution into a broken 
translational symmetry charge density wave state. The 
density profile provides
a way to visualize 
the nature of the state.
At $U=12t$, the region of maximum density tends to
saturate at $\rho=1$,  
while ``stripes'' appear at the boundaries which separate 
AF spin domains with a $\pi$ phase shift.
This is consistent with 
density matrix renormalization group results \cite{Stripe} 
of stripe states 
in Hubbard ladders for $U\gtrsim 8t$.
The real-space characteristic 
length of the charge correlation is $1/2$ that of the spin correlation, 
as Fig.~\ref{fig:large-U} shows.
These results suggest that,
at intermediate $U$, holes are in a uniform
``liquid'' state with no long-range correlation, while at the large $U$ limit
they enter a ``Wigner-crystal'' state forming stripes.

The difficulty in treating the Hubbard model 
arises from the multiple competing energy scales 
separated by tiny differences. The system can fall into one phase or the other
due to a small bias in the calculation or in simulation boundary condition. 
The challenge for numerical calculations is to minimize 
the effect of such biases (intrinsic accuracy, system size, etc). 
Our calculations
reach much larger systems than possible otherwise.
In this work, we have carefully removed  biases other than
the effect of the constrained path approximation. 
To address the latter, we have used
trial wave functions
with opposite properties (uniform FE vs.~broken-symmetry UHF) to
 examine the robustness and 
consistency of the results.

To conclude, we have presented numerical results from constrained path 
QMC to characterize 
the magnetic properties in
doped 2D Hubbard model. 
At intermediate interaction strengths $U/t\sim4$, 
the ground state has incommensurate antiferromagnetic  
SDW order with long wavelength modulation. 
The wavelength of the 
SDW and the strength of the  
spin order both decrease with doping, and the state vanishes below a
critical density, when the system enters a paramagnetic 
``liquid'' phase.
In the SDW state there is essentially no charge correlation, 
with the holes
in a wave-like state. 
As $U$ increases, 
accompanying charge correlation develops, with 
the holes becoming localized 
at the nodal positions of the modulating wave.
Thus in the
strong interaction regime ($U> \sim 10t$) the system
evolves into a stripe-like state. 
Many topics remain for
future work, including quantitative aspects of 
these states and their implications 
on superconductivity.

Computations were carried out at ORNL (Jaguar XT4) 
and William \& Mary (CPD and SciClone clusters). 
We thank Jie Xu and Eric Walter for help, 
and D.~Ceperley, R.~M.~Martin, S.~Sorella, and S.~R.~White for useful comments.
This work is support by ARO (56693-PH) and NSF (DMR-0535592).

\end{document}